\documentclass[preprint]{ptptex}

\title{%        %You can use \\ for explicit line-break.
Eliminating ultraviolet divergence in quantum field theory through use of the Boltzmann factor
}

\author{%       %Use \scshape for the family name.
Kohzo \textsc{Nishida}%
\footnote{E-mail: EZF01671@nifty.com} 
}

\inst{%     %Affiliation, neglected when [addenda] or [errata].
Department of Physics, Kyoto Sangyo University, Kyoto 603-8555, Japan 
}

\abst{%         %This abstract is neglected when [addenda] or [errata].
The need for a cutoff in the Lamb shift calculation suggests that high-energy virtual photons do not interact with real particles.
In this paper, we assume that the creation of high-energy virtual particles is suppressed by a Boltzmann factor.
As a result, the Coulomb potential is modified, and the zero-point energy density and one-particle-irreducible self-energy of the scalar field are finite.
}

\usepackage[]{graphicx}
\begin{document}

\maketitle

\section{Introduction}
In quantum field theory, cutoffs are often introduced when calculating physical quantities.
Let us show this in the Lamb shift\cite{1,2,3,4,5} calculation.
The Lamb shift $\Delta E$ between the $2s_{1/2}$ and $2p_{1/2}$ levels is known to be\cite{6}
\begin{equation}
\label{eq:1}
\Delta E = \frac{\alpha^5 m_e}{6\pi} \int^{m_e}_{1/a_0} \frac{d p}{p}
= \frac{\alpha^5 m_e}{6\pi}  \ln \frac{1}{\alpha},
\end{equation}
where $m_e$ is the mass of the electron, $\alpha$ is the fine structure constant, and $a_0=1/(m_e\alpha)$ is the Bohr radius.
The energy integral of the virtual photon, $\int d p/p$, is stops counting the photons when their wavelength gets bigger than the size of the atom, $a_0$.  
On the short wavelength side, this integral stop counting the photons when their wavelength gets shorter than the Compton wavelength. 
That is, the cutoff $m_e$ is introduced in the Lamb shift calculation.

Renormalization by cutoff is also an operation that does not count virtual particles with energy larger than the cutoff.
Thus, the agreement between the experimental value and the theoretical value introducing the cutoff suggests that 
high-energy virtual photons do not virtually affect real particles.

It is known that equation (\ref{eq:1}) with a cutoff $m_e$ can be approximated to an integral with an infinite range by using a suppression factor $e^{-p/m_e}$ as follows
\begin{equation}
\label{eq:2}
\Delta E \simeq \frac{\alpha^5 m_e}{6\pi} \int^{\infty}_{1/a_0} \frac{d p}{p} e^{-p/m_e}
\simeq  \frac{\alpha^5 m_e}{6\pi}  \left ( \ln \frac{1}{ \alpha} -\gamma + \pi \alpha \right ),
\end{equation}
where we used the integral formula
\begin{equation}
\label{eq:3}
\int^{\infty}_{x} \frac{e^{-p} }{p}d p = \Gamma(0, x) = (-\log(x) -\gamma) + x-\frac{x^2}{4} + \cdots,
\end{equation}
where $\Gamma(s, x)$ is the upper incomplete gamma function, and 
$\gamma=0.57721\cdots$ is  Euler's constant.
Furthermore, we can approximate the integrals
\begin{equation}
\label{eq:4}
I = \int^{\Lambda}_0 p^n dk = \frac{1}{n+1}\Lambda^{n+1}
\,\,\,\,\,\,
(n=0,1,2\cdots)
\end{equation}
in the loop integral as follows when $n$ is small:
\begin{equation}
\label{eq:5}
I \simeq \int^{\infty}_0 p^n e^{-p/\Lambda} d p = n! \Lambda^{n+1}.
\end{equation}
Thus, in general, we can approximate an integral with a cutoff $\Lambda$ as an integral with an infinite integral range multiplied by the suppression factor $e^{-p/\Lambda}$.

The suppression factor $e^{-p/\Lambda}$ has the same form as a Boltzmann factor $e^{-\beta E}$.
If the suppressor is a Boltzmann factor, it means that the creation of virtual particles follows a canonical distribution.
This study aims to investigate how the Boltzmann factor affects quantum field theory.

\section{Modified Coulomb potential}
The momentum representation of the Coulomb potential is
\begin{equation}
\label{eq:6}
V(r)
=
-\frac{Ze^2}{(2\pi)^3}
 \int^{\infty}_{-\infty} d^3 p
 e^{-i \boldsymbol{p}\cdot\boldsymbol{r} }  
\frac{1}{|\boldsymbol{p}|^2},
\end{equation}
where $Z$ is the atomic number.
This potential includes the contribution of high-energy virtual photons.
If the creation of the high-energy virtual photons is suppressed by the Boltzmann factor, equation (\ref{eq:6}) should be modified as follows\cite{7}:
\begin{equation}
\label{eq:7}
V(r)
=
-\frac{Ze^2}{(2\pi)^3}
 \int^{\infty}_{-\infty} d^3 p
 e^{-i \boldsymbol{p}\cdot\boldsymbol{r} }  
\frac{1}{|\boldsymbol{p}|^2}
e^{-|\boldsymbol{p}|/m_c },
\end{equation}
where $m_c$ is a constant with mass as a dimension.
This polar coordinate expression is 
\begin{eqnarray}
\label{eq:8}
V(r)
&=&
-\frac{Ze^2}{(2\pi)^3}
\int_0^\infty dp \int_0^\pi d \theta \int_0^{2\pi} d\phi p^2 \sin \theta e^{-ipr \cos \theta} 
\frac{1}{p^2}
e^{-p/m_c }
\nonumber \\
&=&
-\frac{Ze^2}{(2\pi)^3}
\frac{4\pi}{r}
\int_0^\infty dp \frac{\sin pr}{p} 
e^{-p/m_c },
\end{eqnarray}
where $p=|\boldsymbol{p}|$.
Using the integral formula
\begin{equation}
\label{eq:9}
\int_0^\infty  dx \frac{\sin rx}{x} 
e^{-x/m }
=
\arctan (mr),
\end{equation}
we finally obtain the modified Coulomb potential:
\begin{equation}
\label{eq:10}
V(r)
=
-\frac{Ze^2}{2\pi^2 r} \arctan (m_c r).
\end{equation}
Using $\arctan(x) = x-x^3/3+\cdots$ and $\lim_{x \rightarrow \infty} \arctan(x) = \pi/2$, 
 (\ref{eq:10}) can be approximated in each region as follows:
\begin{eqnarray}
\label{eq:11}
V(r)
&=&
\left \{
\begin{array}{ll}
\displaystyle - \frac{Ze^2 m_c}{2\pi^2} + \frac{Ze^2m_c^3}{6 \pi^2} r^2 & (m_c r \ll 1), \\
 & \\
\displaystyle - \frac{Ze^2}{4\pi r}  & (m_c r \gg 1).
\end{array}
\right.
\end{eqnarray}
This potential has an interesting property that it is an ordinary Coulomb potential at long range and becomes finite at $r = 0$.

\section{Modification of quantum field theory}
In this section, we consider a quantum field theory of scalar fields with the Boltzmann factor.
In the Fourier transforms of position and momentum space, we  request that a relativistic Boltzmann factor be multiplied as follows:
\begin{equation}
\label{eq:12}
f(\boldsymbol{x}) =  \int_{-\infty}^\infty \frac{d^3 p}{ \sqrt{(2\pi)^3}}  F(\boldsymbol{p}) e^{i \boldsymbol{p}\cdot \boldsymbol{x}}
\times e^{-|\bar{u}_\mu p^\mu(\boldsymbol{p})|/(am_c) }
\end{equation}
for position space and 
\begin{equation}
\label{eq:13}
F(\boldsymbol{p}) =  \int_{-\infty}^\infty \frac{d^3 x}{ \sqrt{(2\pi)^3}}  f(\boldsymbol{x}) e^{-i \boldsymbol{p}\cdot \boldsymbol{x}} 
\times e^{|\bar{u}_\mu p^\mu(\boldsymbol{p})|/(am_c) }
\end{equation}
for momentum space, where $a = 1$ or $2$, 
$m_c$ is a constant with mass as a dimension, and 
$p^\mu(\boldsymbol{p}) \equiv\left( \omega(\boldsymbol{p}),\boldsymbol{p} \right)$.
$\bar{u}_\mu$ is an  average of four-velocities of particles created by the scalar field $\phi$.
$e^{|\bar{u}_\mu p^\mu(\boldsymbol{p})|/(am_c) }$ in (\ref{eq:13}) shifts the origin of distribution function like chemical potential.

The spatial distribution of matter in the universe is homogeneous and isotropic, 
which means that the average  created  particle four-velocities $\bar{u}^\mu$ can be written in any coordinate system as 
\begin{equation}
\label{eq:14}
\bar{u}_\mu = (1,0,0,0),
\end{equation}
because the spatial component is canceled with plus and minus appearing equally.
So in any coordinate system, we can always rewrite the the relativistic Boltzmann factor to
\begin{equation}
\label{eq:15}
e^{-|\bar{u}_\mu p^\mu(\boldsymbol{p})|/(am_c) }
=
e^{-\omega(\boldsymbol{p})/(am_c) }.
\end{equation}

From (\ref{eq:12}) with $a=2$, a scalar field is expanded as follows:
\begin{equation}
\label{eq:16}
\phi(t,x) = \int \frac{d^3p}{ \sqrt{(2\pi)^3 2\omega(\boldsymbol{p})}} \{
 a(\boldsymbol{p})e^{-ipx} 
+
a^\dagger(\boldsymbol{p})e^{ipx}
\}
e^{-|u_\mu p^\mu(\boldsymbol{p})|/(2m_c) }.
\end{equation}
The Lagrangian density of the simple free scalar field is given by
\begin{equation}
\label{eq:17}
{\cal L} = \frac{1}{2}\partial_\mu \phi(x) \partial^\mu \phi(x) - \frac{1}{2}m^2(\phi(x))^2.
\end{equation}
The Hamiltonian is given by
\begin{equation}
\label{eq:18}
H = \int d^3 x \{  (\dot{\phi}(x))^2 - {\cal L} \}.
\end{equation}
Substituting (\ref{eq:16}) into the Hamiltonian (\ref{eq:18}), we have
\begin{eqnarray}
\label{eq:19}
H
&=& 
\int d^3p  
\omega(\boldsymbol{p})
\left\{
n(\boldsymbol{p})
+\frac{1}{2} \frac{V_\infty}{ (2\pi)^3}
\right\}
e^{-|\bar{u}_{\mu} p^\mu(\boldsymbol{p})|/m_c },
\nonumber \\
\end{eqnarray}
where $V_\infty \equiv \int d^3x$ and $n(\boldsymbol{p}) = a^\dagger (\boldsymbol{p}) a(\boldsymbol{p})$ is a number operator with momentum $\boldsymbol{p}$. 
Therefore, the zero-point energy density is calculated as
\begin{eqnarray}
\label{eq:20}
\frac{<0|H|0>}{V_\infty} 
&=& 
\int_{-\infty}^{\infty} \frac{d^3p }{ (2\pi)^3} 
\frac{1}{2}\omega(\boldsymbol{p})
e^{-|\bar{u}_\mu p^\mu(\boldsymbol{p})|/m_c } \nonumber \\
&=& 
\int_{-\infty}^{\infty} \frac{d^3p }{ (2\pi)^3}  \frac{1}{2}\omega(\boldsymbol{p})
 e^{-\omega(\boldsymbol{p})/m_c } \nonumber \\
&=& 
\frac{m_c^4  }{(2\pi)^2} \int_{A}^{\infty}   y^2   \sqrt{y^2 -A^2 } e^{-y }  dy  \,\,\,\,\,\,(y^2=\boldsymbol{p}^2/m_c^2+A^2, \,\, A=m/m_c) \nonumber \\
&=& 
\frac{m_c^4   }{2(2\pi)^3} \left \{ \left(\frac{m}{m_c} \right)^3 K_3\left(\frac{m}{m_c} \right) -\left(\frac{m}{m_c} \right)^2 K_2 \left(\frac{m}{m_c} \right) \right\} \nonumber \\
&\leq& 
\frac{m_c^4 }{(2\pi)^2} \int_{A}^{\infty}   y^3  e^{-y }  dy  \nonumber \\
&=& 
\frac{m_c^4 e^{-m/m_c }  }{(2\pi)^2}\left \{ \left(\frac{m}{m_c} \right)^3 +3\left (\frac{m}{m_c} \right )^2 +6\frac{m}{m_c} +6 \right\},
\end{eqnarray}
where we used (\ref{eq:14}). The equal sign can be used only when $m =0$.
$K_2(x)$ and $K_3(x)$ are the modified Bessel function of the second kind.
We obtained  a finite zero-point energy density.

Using (\ref{eq:16}), the Feynman propagator function is calculated,  
\begin{eqnarray}
\label{eq:21}
\lefteqn{ \Delta_F(x,y) } \nonumber \\
&=&
 -i<0|T(\phi(x)\phi(y))|0> \nonumber \\
&=& 
-i\int \frac{d^3 p}{\sqrt{(2\pi)^3 2\omega(\boldsymbol{p})}} \int \frac{d^3 q}{\sqrt{(2\pi)^3 2\omega(\mbox{\boldmath $q$})}} 
e^{-|\bar{u}_\mu p^\mu(\boldsymbol{p})| /(2m_c) }
e^{-|\bar{u}_\mu q^\mu(\boldsymbol{q})| /(2m_c) }
 \nonumber \\
&& 
\times[ <0|\theta(x^0-y^0) a(\boldsymbol{p}) e^{-ipx} a^{\dagger}(\mbox{\boldmath $q$}) e^{iqy}
+\theta(y^0-x^0) a(\mbox{\boldmath $q$}) e^{-iqy} a^{\dagger}(\boldsymbol{p}) e^{ipx} |0>] \nonumber \\
&=&
 -i\int \frac{d^3 p}{(2\pi)^3 2\omega(\boldsymbol{p})} 
e^{-|\bar{u}_\mu p^\mu(\boldsymbol{p})| /m_c }
\nonumber \\
&&
\times \{ \theta(x^0-y^0) e^{-ip(x-y)} + \theta(y^0-x^0) e^{ip(x-y)} \} |_{p^{0} = \omega(\boldsymbol{p}) },
\end{eqnarray}
which can be rewritten as
\begin{eqnarray}
\label{eq:22}
\Delta_F(x,y)
&=&
 \int \frac{d^4 p}{(2\pi)^4} e^{-ip(x-y)} \frac{1}{p^2 -m^2 + i\epsilon} 
e^{-|\bar{u}_\mu p^\mu(\boldsymbol{p})| /m_c }
\end{eqnarray}
using $\omega(-\boldsymbol{p}) = \omega(\boldsymbol{p})$.

Consider the Feynman propagator function in momentum-space:
\begin{eqnarray}
\label{eq:23}
i\Delta_F(p) 
&\equiv& 
\int d^4 x e^{ipx} i\Delta_F(x,0) 
e^{|\bar{u}_\mu p^\mu(\boldsymbol{p})|/m_c }
\nonumber \\
&=& 
 \int \frac{d^4 q}{(2\pi)^4}  \frac{i}{q^2 -m^2 + i\epsilon} 
e^{-|\bar{u}_\mu q^\mu(\boldsymbol{q})|/m_c } (2\pi)^4 \delta^4(q-p)
e^{|\bar{u}_\mu p^\mu(\boldsymbol{p})| /m_c }
 \nonumber \\
&=&
\frac{i}{ p^2 -m^2 + i\varepsilon}.
\end{eqnarray}
Since (\ref{eq:23}) is the Fourier transform of $\Delta_F(x,0)$, we used the modified Fourier transform (\ref{eq:13}) with $a=1$.

Equation (\ref{eq:23}) picking out the specific energy means that the origin of canonical distribution is shifted by $u_\mu p^\mu(\boldsymbol{p})$.
Similarly, we define the one-particle state with specific momentum $\boldsymbol{p}$ as
\begin{equation}
\label{eq:24}
|\mbox{\boldmath $p$}> \equiv e^{|u_\mu p^\mu(\boldsymbol{p})|/(2m_c) } a^{\dagger}(\mbox{\boldmath $p$})  |0>.
\end{equation}
Then, we find that 
the amplitude associated with an external leg with $\boldsymbol{p}$ is 
\begin{eqnarray}
\label{eq:25}
<0| \phi(x) |\boldsymbol{p}> 
 &=&
 \frac{ e^{-ipx} } { \sqrt{ (2\pi)^3 2\omega(\boldsymbol{p}) } } .
\end{eqnarray}
Eventually, from (\ref{eq:23}) and (\ref{eq:25}), 
we find that the Boltzmann factor $e^{-|\bar{u}_\mu p^\mu(\boldsymbol{p})|/m_c } $ 
has no effect on the Feynman rules at the tree level (zero-loop).

Next, consider the case where there is an interaction term ${\cal L}_{int} = -\frac{\lambda}{4!}\phi^4$.
For a first-order two-point function, we have
\begin{eqnarray}
\label{eq:26}
\lefteqn{G^{(2)}_1(x_1, x_2)} \nonumber \\
&\equiv&
\frac{-i\lambda}{2!} \int d^4 y [ i\Delta_F(x_1,y) ] [ i\Delta_F(y,y) ]  [ i\Delta_F(x_2,y) ] \nonumber \\
&=&
 \int d^4 y 
\int \frac{d^4 p_1}{(2\pi)^4}  \frac{ i e^{-ip_1(x_1-y)}}{p_1^2 -m^2 + i\epsilon}
e^{-|\bar{u}_\mu p_1^\mu(\boldsymbol{p}_1)|/m_c } 
\times
\frac{-i\lambda}{2!}
\int \frac{d^4 l }{(2\pi)^4}  \frac{ i e^{-|\bar{u}_\mu l^\mu(\boldsymbol{l})|/m_c } }{l^2 -m^2 + i\epsilon} 
\nonumber \\
& &
\times
\int \frac{d^4 p_2}{(2\pi)^4}  \frac{ i e^{-ip_2(x_2-y)} }{p_2^2 -m^2 + i\epsilon} 
e^{-|\bar{u}_\mu p_2^\mu(\boldsymbol{p}_2)|/m_c } 
\nonumber \\
&=&
\int \frac{d^4 p_1}{(2\pi)^4}  \frac{ i e^{-ip_1x_1}}{p_1^2 -m^2 + i\epsilon}
e^{-|\bar{u}_\mu p_1^\mu(\boldsymbol{p}_1)|/m_c } 
\times
\frac{-i\lambda}{2!}  
\int \frac{d^4 l }{(2\pi)^4}  \frac{ i e^{-|\bar{u}_\mu l^\mu(\boldsymbol{l})|/m_c } }{l^2 -m^2 + i\epsilon} 
\nonumber \\
& &
\times
\int \frac{d^4 p_2}{(2\pi)^4}  \frac{ i e^{-ip_2x_2} }{p_2^2 -m^2 + i\epsilon} 
e^{-|\bar{u}_\mu p_2^\mu(\boldsymbol{p}_2)|/m_c } 
(2\pi)^4 \delta^4(p_2+p_1)
\nonumber \\
&=&
\int \frac{d^4 p_1}{(2\pi)^4} 
\frac{ i e^{-ip_1(x_1-x_2)} }{p_1^2 -m^2 + i\epsilon}
e^{-|\bar{u}_\mu p_1^\mu(\boldsymbol{p}_1)|/m_c } 
\times \frac{-i\lambda}{2!} \int \frac{d^4 l }{(2\pi)^4} 
\frac{ i e^{-|\bar{u}_\mu l^\mu(\boldsymbol{l})|/m_c } }{l^2 -m^2 + i\epsilon} 
 \nonumber \\
&&
\times
\frac{ i  }{p_1^2 -m^2 + i\epsilon}
e^{-|\bar{u}_\mu p_1^\mu(\boldsymbol{p}_1)|/m_c }.
\end{eqnarray}

Using the modified Fourier transform (\ref{eq:13}) with $a=1$, we obtain the two-point Green's function in momentum-space 
\begin{eqnarray}
\label{eq:27}
i\Delta'_F (p) 
&=& 
\int d^4 x e^{ipx}
\{ 
i\Delta_F(x,0) + G^{(2)}_1(x, 0) +\cdots
 \} 
e^{|\bar{u}_\mu p^\mu(\boldsymbol{p})|/m_c }
\nonumber \\
&=& 
\frac{ i  }{ p^2 -m^2 +i\epsilon} 
+  
\frac{ i  }{ p^2 -m^2 +i\epsilon} 
\{ -i\Sigma(p^2) \} 
\frac{ i e^{-|\bar{u}_\mu p^\mu(\boldsymbol{p})|/m_c } }{ p^2 -m^2 +i\epsilon} 
 + \cdots 
\nonumber \\
&=& 
\frac{ i  }{ p^2 -m^2 - \Sigma(p^2) e^{-|\bar{u}_\mu p^\mu(\boldsymbol{p})|/m_c }   + i\epsilon},
\end{eqnarray}
where $-i\Sigma(p^2)$ is the one-particle-irreducible self-energy function
\begin{eqnarray}
\label{eq:28}
-i\Sigma(p^2)
&=&
\frac{-i\lambda}{2} \int \frac{d^4 l }{ (2\pi)^4 } \frac{i } { l^2 -m^2 + i\epsilon } 
e^{-|\bar{u}_\mu l^\mu(\boldsymbol{l})|/m_c }.
\end{eqnarray}
Let us calculate $\Sigma(p^2) $ in (\ref{eq:28}). We have
\begin{eqnarray}
\label{eq:29}
\Sigma(p^2)
&=&
 \frac{\lambda}{2}  \int \frac{d^4 l }{ (2\pi)^4 } \frac{i  } { l^2 -m^2 + i\epsilon } 
e^{-|\bar{u}^\mu l_\mu(\boldsymbol{l})|/m_c } 
\nonumber \\
&=&
 \frac{i \lambda}{2}  \int \frac{d^3 l }{ (2\pi)^4 }  
e^{-|\bar{u}^\mu l_\mu(\boldsymbol{l})|/m_c } 
\int_C d l^0   \frac{1  } { (l^0-\omega + i\epsilon)(l^0+\omega - i\epsilon) } 
\nonumber \\
&=&
 \frac{i \lambda}{2}  \int \frac{d^3 l }{ (2\pi)^4 }    
e^{-|\bar{u}^\mu l_\mu(\boldsymbol{l})|/m_c } 
\left.
\frac{ 2\pi i  } { l^0-\omega + i\epsilon} \right|_{l^0 = -\omega+i\epsilon}
 \nonumber \\
&=&
  \frac{\lambda}{2} \int_{-\infty}^{\infty} \frac{d^3 l}{(2\pi)^3 2\omega(\mbox{\boldmath $l$})} 
e^{-|\bar{u}^\mu l_\mu(\boldsymbol{l})|/m_c }.
\end{eqnarray}
Again, using $\bar{u}^\mu = (1,0,0,0)$, 
(\ref{eq:29}) is calculated as
\begin{eqnarray}
\label{eq:30}
\Sigma(p^2)
&=&
  \frac{\lambda}{2} \int_{-\infty}^{\infty} \frac{d^3 l}{(2\pi)^3 2\omega(\mbox{\boldmath $l$})} 
e^{-\omega(\boldsymbol{l})/m_c }
 \nonumber \\
&=&
 \frac{ \lambda m_c^2 }{ 4(2\pi)^3}4\pi \int_A^\infty \sqrt{ y^2 -A^2} e^{-x } dy  \,\,\,\,\,\, (A=m/m_c)  \nonumber \\
&=&
\frac{ \lambda m_c^2  }{ 8\pi^2} \frac{m}{m_c} K_1\left(\frac{m}{m_c} \right) \nonumber \\
&\leq&
\frac{\lambda m_c^2 }{ 8\pi^2} \int_A^\infty x e^{-y } dy  \,\,\,\,\,\, \nonumber \\
&=&
 \frac{\lambda m_c^2  e^{-m/m_c} }{ 8\pi^2} \left( \frac{m}{m_c} +1 \right),
\end{eqnarray}
where the equal sign can be used only when $m =0$. 
$K_1(x)$ is is the modified Bessel function of the second kind.
We obtained  a finite self-energy.

Finally, let us give one of the four-point vertex correction
 to see an example of the Boltzmann factor:
\begin{eqnarray}
\label{eq:31}
\lefteqn{ i\tilde{ \Gamma}^{(4)} ( (p_1+p_2)^2) } \nonumber \\  
&=&
\frac{(-i \lambda)^2}{2} \int \frac{ d^4 l }{(2\pi)^4}
\frac{ i e^{-|\bar{u}^\mu l_\mu(\boldsymbol{l})|/m_c }  } { l^2 -m^2 + i\epsilon } 
\times \frac{ i e^{- | \bar{u}^\mu  ( p_1^\mu(\boldsymbol{p}_1)+p_2^\mu(\boldsymbol{p}_2) - l^\mu(\boldsymbol{l}) )|/m_c } } 
{ \{ (p_1+p_2) - l  \} ^2 -m^2 + i\epsilon } .
\nonumber \\
\end{eqnarray}

\section{Conclusion}
In the first section, we have shown that the introduction of cutoff into theory is equivalent to the introduction of the Boltzmann factors.
Next, we derived a non-relativistic Coulomb potential in which high-energy virtual photons are suppressed by the Boltzmann factor.
Furthermore, we tried to introduce the relativistic Boltzmann factor into quantum field theory, and 
demonstrated that the factor results in finite values for the zero-point energy density and one-particle-irreducible self-energy of the scalar field.

\end{document}